\documentclass[a4paper,10pt,eqno]{article}
\usepackage{latexsym,amssymb,amsfonts,amsmath}
\usepackage{amsthm}
\usepackage[dvips]{graphicx}
\usepackage{color}
\newtheorem{thm}{Theorem}[section]
\newtheorem*{thm*}{Theorem}
\newtheorem{lem}[thm]{Lemma}
\newtheorem*{lem*}{Lemma}

\newtheorem*{cor*}{Corollary}

\newcommand{\ZM}{\mathbb{Z}}

\newcommand{\NM}{\mathbb{N}}
\newcommand{\AM}{\mathbb{A}}
\newcommand{\CM}{\mathbb{C}}

\newcommand{\QM}{\mathbb{Q}}

\newcommand{\HC}{\mathcal{H}}

\newcommand{\ket}[1]{|#1\rangle}
\newcommand{\bra}[1]{\langle #1|}

\newcommand{\lr}[1]{\left( #1 \right)}

\newcommand{\bmat}[1]{\begin{bmatrix} #1 \end{bmatrix}}

\title{{\Large {\bf Periodicity for the 3-state quantum walk on cycles}}}
\author{
{\small Takeshi Kajiwara\footnote{kajiwara-takeshi-rj@ynu.ac.jp},} \quad 
{\small Norio Konno\footnote{konno-norio-bt@ynu.ac.jp},} \quad 
{\small Shohei Koyama\footnote{koyama-shohei-mw@ynu.jp},} \quad 
{\small Kei Saito\footnote{saito-kei-nb@ynu.jp},}
\\
{\scriptsize  Department of Applied Mathematics, Faculty of Engineering Science, Yokohama National University}\\
{\scriptsize \footnotesize\it 79-5 Tokiwadai, Hodogaya, Yokohama, 240-8501, Japan}\\
}
\date{\empty}
\begin{document}
\maketitle

\par\noindent
\begin{small}
\par\noindent
{\bf Abstract}.
Dukes (2014) and Konno, Shimizu, and Takei (2017) studied the periodicity for 2-state quantum walks whose coin operator is the Hadamard matrix on cycle graph $C_N$ with $N$ vertices.
The present paper treats the periodicity for 3-state quantum walks on $C_N$.
Our results follow from a new method based on cyclotomic field.
This method shows a necessary condition for the coin operator of quantum walks to have the finite period.
Moreover, we reveal the period $T_N$ of two kinds of typical quantum walks, the Grover and Fourier walks.
We prove that both walks do not have any finite period except for $N=3$,
in which case $T_3=6$ (Grover), $=12$ (Fourier).

\footnote[0]{
{\it Abbr. title:} Periodicity for 3-state quantum walk on cycles
}
%\footnote[0]{
%{\it AMS 2000 subject classifications: }
%60F05, 81P68
%}
\end{small}

\setcounter{equation}{0}

\section{Introduction}
The discrete time quantum walk is defined as a quantum counterpart of the classical random walk \cite{QW1, QW2}.
It is known that theory of quantum walks is useful for developing new quantum algorithm.
As other applications, quantum walks have been studied from various research fields, e.g., topological insulator, quantum information science.
Some reviews and books for the quantum walk are \cite{review1,review2,review3,review4}.
In the present paper, we treat 3-state quantum walks on the cycle graph $C_N$ with $N$ vertices.
Especially, we focus on the property of periodicity.
The periodicity of the quantum walk is widely studied with some applications \cite{period1, period2, period3}.
For example, one of our motivations of the study on the periodicity of quantum walks is to characterise graphs \cite{period7, period9}.
Moreover, the periodicity is clarified for some typical graphs, e.g., complete graphs, the generalized Bethe trees, and cycles \cite{period4, period5, period6, period8}. 
We should remark that the 3-state quantum walk on cycle is regarded as the quantum walk on cycles with self-loops.

From now on, we present definition of 3-state quantum walks on $C_N$ introduced by  Sadowski et al.\cite{lively}.
Furthermore, this model is given by adding a periodic boundary condition to 3-state quantum walks on $\ZM$ which is a typical walk with localization \cite{1d3s1, 1d3s2}, where $\ZM$ means the set of integers.
Firstly, we consider a quantum walk on the Hilbert space $\HC = V(C_N)\otimes \CM^3$, 
where $V(C_N)=\{0,1,\cdots ,N-1\}$ is the vertex set of $C_N\ (N\geq 2)$.
The quantum walker has three kinds of chirality states, $\leftarrow$, $\bullet$, and $\rightarrow$.
Each state means the direction of motion of the walker.
Here, we correspond each state to the vectors as follows.
\begin{align*}
\ket{\leftarrow}=\bmat{1 \\ 0 \\ 0},\quad
\ket{\bullet}=\bmat{0 \\ 1 \\ 0},\quad
\ket{\rightarrow}=\bmat{0 \\ 0 \\ 1}.
\end{align*}
Weights $P,Q$, and $R$ represent the walker hops to adjacent vertices or stays the same vertex.
They are given by a division of the coin operator $C\in{\rm U}(3)$, respectively:
\begin{align*}
P=\ket{\leftarrow}\bra{\leftarrow}C,\quad
R=\ket{\bullet}\bra{\bullet}C,\quad
Q=\ket{\rightarrow}\bra{\rightarrow}C,
\end{align*}
where ${\rm U}(n)$ means the set of $n\times n$ unitary matrices.
Here, we note that above definitions of $P,R,$ and $Q$ are moving shift type which is often used for quantum walks on some fixed graphs, e.g., $n$-dimensional lattice, tree graphs.
On the other hand, the flip-flop shift is used to define quantum walks on general graphs.
The results in this paper are only for the moving shift type.
However, we should remark that similar results can be obtained in flip-flop shift.
 
For the initial state $\Psi_0\in\HC$, the time evolution is defined as $\Psi_t=U_N^t\Psi_0$, where the time evolution operator $U_N$ is given as follows.
\begin{align*}
{U_N}=
\bmat{
R & P & O & \cdots & O & Q \\
Q & R & P & \ddots & O & O \\
O & Q & R & \ddots & \ddots & \vdots \\
\vdots & \ddots & \ddots & \ddots & \ddots & \vdots \\
O & O & \ddots & \ddots & R & P \\
P & O & \cdots & \cdots & Q & R
}\in{\rm U}(3N)\quad (N\neq 2).
\end{align*}
For $N=2$, we define
\begin{align*}
U_2=\bmat{
R & P+Q \\ P+Q & R
}\in{\rm U}(6).
\end{align*} 
Secondly, the Fourier transform of the quantum state is defined as $\hat\Psi_t(k)=\sum_{x=0}^{N-1}e^{-\frac{2\pi k}{N}i}\Psi_t(x)\in\CM^3 \ (k=0,1,\ldots , N-1)$, where
\begin{align*}
\Psi_t={}^t \bmat{
\Psi_t(0),\ \Psi_t(1),\ \cdots ,\ \Psi_t(N-2),\ \Psi_t(N-1)
}\in\HC.
\end{align*}
Here, $t$ means the transpose operation.
Then, $\hat{U}(k)={\rm diag}(e^{\frac{2\pi k}{N}i},\ 1,\ e^{-\frac{2\pi k}{N}i})C\in{\rm U}(3)$ gives the time evolution for the quantum state on the Fourier space, i.e., $\hat\Psi_t(k)=\lr{\hat{U}(k)}^t\hat\Psi_0(k)$.
Hence, we can check ${\rm Spec}(U_N)=\bigcup_{k=0}^{N-1}{\rm Spec}(\hat{U}(k))$, where ${\rm Spec}(\cdot)$ denotes the set of eigenvalues.

Next, we put
\begin{align*}
\mathcal{N}=\{n\in \NM\ :\ \lr{U_N}^T=I_{3N}\},
\end{align*}
where $\NM$ is the set of natural numbers and $I_n$ is the $n\times n$ identity matrix.
If $\mathcal{N}\neq \emptyset$, the period of the quantum walk $T_N$ is defined by $\min\mathcal{N}$.
If $\mathcal{N}=\emptyset$, then $T_N=\infty$ and we say the quantum walk is not periodic. 
Thus, the following lemma is a basic method to distinguish the quantum walk is periodic or not.
\begin{lem}
\label{lem1}
For $T\in\NM$, the following (1) to (3) are equivalent.
\begin{align*}
&\text{(1)}\quad U_N^T = I_{3N}.
\\
&\text{(2)}\quad {}^\forall\lambda\in{\rm Spec}(U_N),\ \lambda^T=1.
\\
&\text{(3)}\quad {}^\forall\lambda(k)\in{\rm Spec}(\hat{U}(k)),\ \lambda(k)^T=1\quad (k=0,1,\ldots , N-1).
\end{align*}
\end{lem}
We should remark that $\ZM[\zeta_N]=\AM\cap \QM[\zeta_N]$, where $\ZM[\zeta_N]$ is the ring of integers of the $n$-th cyclotomic field, $\AM$ is the set of algebraic integers, and $\QM[\zeta_N]$ is the $n$-th cyclotomic field.
The following result is easily obtained by definitions of $\AM$ and $\QM[\zeta_N]$.
Moreover, $\zeta_n$ is a primitive $n$-th root of unity, i.e., $\zeta_n=e^{\frac{2\pi}{n}i}$. 
\begin{lem}
\label{lem2}
For $T\in\NM$, if $U_N^T=I_{3N}$, then the following relation holds.
\begin{align*}
\lambda_1(k)+\lambda_2(k)+\lambda_3(k)\in\AM\cap\QM[\zeta_T]=\ZM[\zeta_T]
\qquad (k=0,1,\ldots ,N-1)
,
\end{align*}
where $\lambda_1(k),\lambda_2(k),\lambda_3(k)\in{\rm Spec}(\hat{U}(k))$.
\end{lem}

\section{Results}
In this section, we will present some results of the periodicity of the quantum walk.
\subsection{Necessary condition for the coin operator} 
We will show a necessary condition for the coin operator to have the finite period.
\begin{thm}
\label{Thm2}
If $c_{11}\not\in\frac{1}{N}\ZM[\zeta_{{\rm lcm}(N,T)}]$ or $c_{22}\not\in\frac{1}{N}\ZM[\zeta_{T}]$ or $c_{33}\not\in\frac{1}{N}\ZM[\zeta_{{\rm lcm}(N,T)}]$ is satisfied, then $U_N^T\neq I_{3N}$ for any $T\in\NM$,
where $C=(c_{ij})_{i,j=1,2,3}$, ${\rm lcm}(\cdot, \cdot)$ is the least common multiple.
\end{thm}
\noindent
{\bf Proof:}\quad 
Assume that $U_N^T=I_{3N}$. 
From Lemma \ref{lem2}, we see that
\begin{align}
\label{siki7}
\lambda_1(k)+\lambda_2(k)+\lambda_3(k)\in\ZM[\zeta_T].
\end{align}
By definition of $\hat{U}(k)$, we have
\begin{align}
\label{siki8}
{\rm tr}(\hat{U}(k))=e^{\frac{2\pi k}{N}i}c_{11} + c_{22} + e^{-\frac{2\pi k}{N}i}c_{33}.
\end{align}
Combining (\ref{siki7}) with (\ref{siki8}) gives
\begin{align}
\label{siki1}
e^{\frac{2\pi k}{N}i}c_{11} + c_{22} + e^{-\frac{2\pi k}{N}i}c_{33} \in \ZM[\zeta_T].
\end{align}
It follows from (\ref{siki1}) and $e^{\pm\frac{2\pi k}{N}i}\ZM[\zeta_T]\subset \ZM[\zeta_{{\rm lcm}(N,T)}]$ that
\begin{align*}
&\sum_{k=0}^{N-1}\lr{e^{\frac{2\pi k}{N}i}c_{11} + c_{22} + e^{-\frac{2\pi k}{N}i}c_{33}}
=Nc_{22}\in \ZM[\zeta_T],
\\
&\sum_{k=0}^{N-1}e^{-\frac{2\pi k}{N}i}\lr{e^{\frac{2\pi k}{N}i}c_{11} + c_{22} + e^{-\frac{2\pi k}{N}i}c_{33}}
=Nc_{11}\in \ZM[\zeta_{{\rm lcm}(N,T)}],
\\
&\sum_{k=0}^{N-1}e^{\frac{2\pi k}{N}i}\lr{e^{\frac{2\pi k}{N}i}c_{11} + c_{22} + e^{-\frac{2\pi k}{N}i}c_{33}}
=Nc_{33}\in \ZM[\zeta_{{\rm lcm}(N,T)}].
\end{align*}
Therefore, if $U_N^T\ = I_{3N}$, then $c_{11}\in\frac{1}{N}\ZM[\zeta_{{\rm lcm}(N,T)}]$, $c_{22}\in\frac{1}{N}\ZM[\zeta_{T}]$, and $c_{33}\in\frac{1}{N}\ZM[\zeta_{{\rm lcm}(N,T)}]$.
The desired conclusion is given.

\subsection{Periodicity for the typical quantum walks}
We reveal the periodicity of two kinds of typical quantum walks, the Grover and Fourier walks.
\subsubsection{The Grover walk}
The Grover walk is determined by the coin operator as the Grover matrix, i.e.,
\begin{align*}
C=G(3)=\frac{1}{3}\bmat{
-1 & 2 & 2\\
2 & -1 & 2\\
2 & 2 & -1
}.
\end{align*}
Here, $G(n)=(g_{i,j})_{i,j=1,2,\ldots ,n}$ is the Grover matrix with size $n$, which is defined by 
\begin{align*}
g_{i,j}=
\begin{cases}
\frac{2}{n} -1\quad &(i=j)
\\
\frac{2}{n} \quad &(i\neq j)
\end{cases}.
\end{align*}
Then, the characteristic polynomial of $\hat{U}(k)$ is 
\begin{align}
\label{siki2}
{\rm det}\lr{\lambda(k)I_3 - \hat{U}(k)}=(\lambda(k) - 1)\lr{\lambda(k)^2 + \frac{4+e^{\frac{2\pi k}{N}i}+e^{-\frac{2\pi k}{N}i}}{3}\lambda(k) + 1}.
\end{align}
The above equation gives ${\rm Spec}(\hat{U}(k))=\{\lambda_1(k), \lambda_2(k), \lambda_3(k) \}$ with $\lambda_2(k)=\overline{\lambda_1(k)}$ and $\lambda_3(k)=1$.
\begin{thm}
\label{Thm3}
For any $N\geq 2$, the period of the Grover walk is as follows.
\begin{align*}
T_N=\begin{cases}
6\quad &(N=3)
\\
\infty\quad &(N\neq 3)
\end{cases}.
\end{align*}
\end{thm}
\noindent
{\bf Proof:}\quad 
Firstly, we prove $N=3$ case.
From (\ref{siki2}), we get the eigenvalues of $\hat{U}(k)$ as follows.
\begin{align*}
\lambda_1(k)=\overline{\lambda_2(k)}=\begin{cases}
-1\quad &(k=0) \\
e^{\frac{2\pi k}{3}i} \quad &(k=1,2)
\end{cases}.
\end{align*}
Hence, Lemma \ref{lem1} gives $T_3={\rm lcm}(2,3)=6$.
Next, we prove $N\neq 3$ case.
Assume that $U_N^T=I_{3N}$.
By Lemma \ref{lem2}, we see that
\begin{align*}
\lambda_1(k)+\lambda_2(k)+\lambda_3(k)\in \AM \cap \QM[\zeta_T].
\end{align*}
Since $\lambda_3(k)=1$, we have
\begin{align}
\label{siki3}
\lambda_1(k)+\lambda_2(k)\in \AM \cap \QM[\zeta_T].
\end{align}
On the other hand, definition of $\QM[\zeta_N]$ and (\ref{siki2}) imply
\begin{align}
\label{siki9}
\lambda_1(k)+\lambda_2(k) = -\frac{4+e^{\frac{2\pi k}{N}i}+e^{-\frac{2\pi k}{N}i}}{3} \in\QM[\zeta_N].
\end{align}
Combining $\ZM[\zeta_N]=\AM\cap\QM[\zeta_N]$ with (\ref{siki3}) and (\ref{siki9}) gives
\begin{align*}
\lambda_1(k)+\lambda_2(k)\in\ZM[\zeta_N].
\end{align*}
Especially, we now focus on $k=1$ case:
\begin{align*}
\lambda_1(1)+\lambda_2(1)=-\frac{4+e^{\frac{2\pi}{N}i}+e^{-\frac{2\pi}{N}i}}{3}\in\ZM[\zeta_N].
\end{align*}
Here, we should remark that
\begin{align}
\label{siki4}
3e^{\frac{2\pi}{N}i}\lr{
-\frac{4+e^{\frac{2\pi}{N}i}+e^{-\frac{2\pi}{N}i}}{3}
}
\in 3 \ZM[\zeta_N].
\end{align}
On the other hand, we have
\begin{align}
\label{siki10}
3e^{\frac{2\pi}{N}i}\lr{
-\frac{4+e^{\frac{2\pi}{N}i}+e^{-\frac{2\pi}{N}i}}{3}
}
=
-1-4e^{\frac{2\pi}{N}i}-e^{\frac{4\pi}{N}i}.
\end{align}
Combining (\ref{siki4}) with (\ref{siki10}) implies
\begin{align}
\label{siki11}
-1-4e^{\frac{2\pi}{N}i}-e^{\frac{4\pi}{N}i}=
-\zeta_N^{0}-4\zeta_N^{1}-\zeta_N^2
\in 3 \ZM[\zeta_N].
\end{align}
In general, $x\in\ZM[\zeta_n]$ is uniquely expressed by a linear combination of $\zeta_n^{0}$ to $\zeta_n^{\phi(n)-1}$, i.e.,
\begin{align*}
x=\sum_{j=0}^{\phi(n)-1}z_j \zeta_n^j,
\end{align*} 
where $\{z_j\}$ is a sequence of integer numbers and $\phi(n)$ is Euler's totient function.
If $\phi(N)>2$, then a contradiction occurs in the assumption $U_N^T=I_{3N}$ because the coefficients of $e^{\frac{2\pi\times 0}{n}i}, e^{\frac{2\pi\times 1}{n}i}$ and $e^{\frac{2\pi\times 2}{n}i}$ in (\ref{siki11}) do not belong to $3\ZM$.
In the rest of the proof, we consider $N$ with $\phi(N)\leq 2$ and $N\neq 3$, i.e., $N=2,4,$ and $6$ cases.
For three cases, it follows from (\ref{siki11}) and the following easily obtained results (i) to (iii) that a contradiction occurs in the assumption $U_N^T=I_{3N}$. Therefore, the desired conclusion is given.
\begin{align*}
\text{(i)}\quad &N=2 \text{ case : }-\zeta_2^0-4\zeta_2^1-\zeta_2^2=\hspace{0.27cm} 2\zeta_2^0\ \not\in 3 \ZM[\zeta_2]
\\
\text{(ii)}\quad &N=4 \text{ case : }-\zeta_4^0-4\zeta_4^1-\zeta_4^2=-4\zeta_4^1\ \not\in 3 \ZM[\zeta_4]
\\
\text{(iii)}\quad &N=6 \text{ case : }-\zeta_6^0-4\zeta_6^1-\zeta_6^2=-5\zeta_6^1\ \not\in 3 \ZM[\zeta_6]
\end{align*}
%\begin{align*}
%\text{(i)}\quad &N=2 \text{ case : }-1-4e^{\frac{2\pi}{2}i}-e^{\frac{4\pi}{2}i}=\hspace{0.27cm} 2e^{\frac{2\pi\times 0}{2}i}\ \not\in 3 \ZM[\zeta_2]
%\\
%\text{(ii)}\quad &N=4 \text{ case : }-1-4e^{\frac{2\pi}{4}i}-e^{\frac{4\pi}{4}i}=-4e^{\frac{2\pi\times 1}{4}i}\ \not\in 3 \ZM[\zeta_4]
%\\
%\text{(iii)}\quad &N=6 \text{ case : }-1-4e^{\frac{2\pi}{6}i}-e^{\frac{4\pi}{6}i}=-5e^{\frac{2\pi\times 1}{6}i}\ \not\in 3 \ZM[\zeta_6]
%\end{align*}

\subsubsection{The Fourier walk}
The Fourier walk is defined by the coin operator as the Fourier matrix, i.e.,
\begin{align*}
C=F(3)=\frac{1}{\sqrt{3}}\bmat{
1 & 1 & 1\\
1 & e^{\frac{2\pi}{3}i} & e^{\frac{4\pi}{3}i}\\
1 & e^{\frac{4\pi}{3}i} & e^{\frac{2\pi}{3}i}
}.
\end{align*}
Here, $F(n)=(e^{\frac{2(u-1)(v-1)\pi}{n}i}/\sqrt{n}\,)_{u,v=1,2,\ldots ,n}$ is the Fourier matrix with size $n$.
Then, the characteristic polynomial of $\hat{U}(k)$ is 
\begin{align}
\label{siki5}
\nonumber
{\rm det}\lr{\lambda(k)I_3 - \hat{U}(k)}=
\lambda(k)^3&-\frac{\sqrt{3}}{3}\lr{
e^{\frac{2\pi k}{N}i} + e^{\lr{\frac{-2\pi k}{N}+\frac{2\pi}{3}}i} + e^{\frac{2\pi}{N}i}
}\lambda(k)^2
\\
&-
\frac{1}{3}\lr{
1+e^{\frac{2\pi k}{N}i}+e^{\lr{\frac{-2\pi k}{N}+\frac{2\pi}{3}}i}
-e^{\frac{2\pi}{3}i} - e^{\lr{\frac{2\pi k}{N}+\frac{2\pi}{3}}i}- e^{\lr{\frac{-2\pi k}{N}+\frac{4\pi}{3}}i}
}\lambda(k)
+i.
\end{align}
As in the case of the Grover walk, the following result on the periodicity for the Fourier walk can be obtained by (\ref{siki5}).
\begin{thm}
\label{Thm4}
For any $N\geq 2$, the period of the Fourier walk is as follows.
\begin{align*}
T_N=\begin{cases}
12\quad &(N=3)
\\
\infty\quad &(N\neq 3)
\end{cases}.
\end{align*}
\end{thm}
\noindent
{\bf Proof:}\quad 
Firstly, we check $N=3$ case.
Then, (\ref{siki5}) implies that
\begin{align*}
(\lambda_1(k), \lambda_2(k), \lambda_3(k))=
\begin{cases}
(e^{\frac{2\pi}{4}i}, 1, -1)\quad &(k=0,1)\\
(e^{\frac{2\pi}{3}i}, e^{\frac{5\pi}{3}i}, e^{\frac{7\pi}{6}i})\quad &(k=2)
\end{cases}.
\end{align*}
Hence, Lemma \ref{lem1} gives $T_3={\rm lcm}(4,6)=12$.
Next, we want to show that $T_N=\infty$ if $N\neq 3$.
As for $N\neq 2$ or $3^n\ (n\in\NM)$ cases, $T_N=\infty$ can be derived from the result given by Saito \cite{period9} in our setting:
\begin{thm}
\label{thmSaito}
{\rm \bf \cite{period9}}\ For $N\neq 3^n\ (N>2)$ with $n\in\NM$, the Fourier walk on $C_N$ is not periodic.
\end{thm}
\noindent
Therefore, we will prove only for $N=2$ and $9$ cases, since ${\rm Spec}(U_{9})\subset{\rm Spec}(U_{3^n})\ (n>2)$ and Lemma \ref{lem1}.
As in the proof of Theorem \ref{Thm3}, we focus on the following $\lambda_1(1)+\lambda_2(1)+\lambda_3(1)$ given by (\ref{siki5}).
\begin{align}
\label{siki6}
\lambda_1(1)+\lambda_2(1)+\lambda_3(1) = \frac{1}{3}\lr{
1+e^{\frac{2\pi}{N}i}+e^{\lr{-\frac{2\pi}{N}+\frac{2\pi}{3}}i}
-e^{\frac{2\pi}{3}i} - e^{\lr{\frac{2\pi}{N}+\frac{2\pi}{3}}i}- e^{\lr{\frac{-2\pi}{N}+\frac{4\pi}{3}}i}
}.
\end{align}
Assume that $U^T_N=I_{3N}$. Then, Lemma \ref{lem2} implies
\begin{align}
\label{siki13}
\lambda_1(1)+\lambda_2(1)+\lambda_3(1)\in\AM\cap\QM[\zeta_T].
\end{align}
By definition of $\QM[\zeta_N]$ and (\ref{siki6}), we obtain
\begin{align}
\label{siki14}
\lambda_1(1)+\lambda_2(1)+\lambda_3(1)\in\QM[\zeta_{{\rm lcm}(3,N)}].
\end{align}
Thus, Combining (\ref{siki13}) with (\ref{siki14}) gives
\begin{align*}
\frac{1}{3}\lr{
1+e^{\frac{2\pi}{N}i}+e^{\lr{-\frac{2\pi}{N}+\frac{2\pi}{3}}i}
-e^{\frac{2\pi}{3}i} - e^{\lr{\frac{2\pi}{N}+\frac{2\pi}{3}}i}- e^{\lr{\frac{-2\pi}{N}+\frac{4\pi}{3}}i}
}
\in\ZM[\zeta_{{\rm lcm}(3,N)}].
\end{align*}
Then, we have
\begin{align}
\label{siki12}
\nonumber
1+e^{\frac{2\pi}{N}i}+e^{\lr{-\frac{2\pi}{N}+\frac{2\pi}{3}}i}
&-e^{\frac{2\pi}{3}i} - e^{\lr{\frac{2\pi}{N}+\frac{2\pi}{3}}i}- e^{\lr{\frac{-2\pi}{N}+\frac{4\pi}{3}}i}
\\
&=\zeta_N^{0}+\zeta_N^{1}+\zeta_3^{1}\zeta_N^{-1}-\zeta_3^{1}-\zeta_3^{1}\zeta_N^{1}-\zeta_3^{2}\zeta_N^{-1}
\in 3\ZM[\zeta_{{\rm lcm}(3,N)}].
\end{align}
From (\ref{siki12}) and the following results (i) and (ii), a contradiction occurs in the assumption $U^T_N=I_{3N}$. 
Hence, desired conclusion is given.
\begin{align*}
\text{(i)}\quad &N=2 \text{ case : }1+e^{\frac{2\pi}{2}i}+e^{\lr{-\frac{2\pi}{2}+\frac{2\pi}{3}}i}
-e^{\frac{2\pi}{3}i} - e^{\lr{\frac{2\pi}{2}+\frac{2\pi}{3}}i}- e^{\lr{\frac{-2\pi}{2}+\frac{4\pi}{3}}i} 
\\
&\hspace{8.3 cm}= \zeta_6^{0}-2\zeta_6^{1} \not\in 3 \ZM[\zeta_{{\rm lcm}(3,2)}]=3 \ZM[\zeta_6]
\\[+9pt]
\text{(ii)}\quad &N=9 \text{ case : }1+e^{\frac{2\pi}{9}i}+e^{\lr{-\frac{2\pi}{9}+\frac{2\pi}{3}}i}
-e^{\frac{2\pi}{3}i} - e^{\lr{\frac{2\pi}{9}+\frac{2\pi}{3}}i}- e^{\lr{\frac{-2\pi}{9}+\frac{4\pi}{3}}i}
\\
&\hspace{5.4 cm}=\zeta_9^{0}+\zeta_9^{1}+\zeta_9^{2}-\zeta_9^{3}-\zeta_9^{4}-\zeta_9^{5}
 \not\in 3 \ZM[\zeta_{{\rm lcm}(3,9)}]=3 \ZM[\zeta_9]
\end{align*}
%\begin{align*}
%\text{(i)}\quad &N=2 \text{ case : }1+e^{\frac{2\pi}{2}i}+e^{\lr{-\frac{2\pi}{2}+\frac{2\pi}{3}}i}
%-e^{\frac{2\pi}{3}i} - e^{\lr{\frac{2\pi}{2}+\frac{2\pi}{3}}i}- e^{\lr{\frac{-2\pi}{2}+\frac{4\pi}{3}}i} 
%\\
%&\hspace{9.3 cm}= e^{\frac{2\pi\times 0}{9}i}-2e^{\frac{2\pi\times 1}{6}i} \not\in 3 \ZM[\zeta_{{\rm lcm}(3,2)}]=3 \ZM[\zeta_6]
%\\[+9pt]
%\text{(ii)}\quad &N=9 \text{ case : }1+e^{\frac{2\pi}{9}i}+e^{\lr{-\frac{2\pi}{9}+\frac{2\pi}{3}}i}
%-e^{\frac{2\pi}{3}i} - e^{\lr{\frac{2\pi}{9}+\frac{2\pi}{3}}i}- e^{\lr{\frac{-2\pi}{9}+\frac{4\pi}{3}}i}
%\\
%&\hspace{4.0 cm}=e^{\frac{2\pi\times 0}{9}i}+e^{\frac{2\pi\times 1}{9}i}+e^{\frac{2\pi\times 2}{9}i}-e^{\frac{2\pi\times 3}{9}i}-e^{\frac{2\pi\times 4}{9}i}-e^{\frac{2\pi\times 5}{9}i}
% \not\in 3 \ZM[\zeta_{{\rm lcm}(3,9)}]=3 \ZM[\zeta_9]
%\end{align*}

\section{Summary and discussion}
In the present paper, we got some results about the periodicity of 3-state quantum walks on cycles by using a method of cyclotomic field.
Especially, we completely determined the periodicity of two kinds of typical quantum walks, the Grover and Fourier walks, in Theorems {\ref{Thm3}} and {\ref{Thm4}}, respectively.
We should remark that there are two kinds of typical shifts for quantum walks, i.e., moving and flip-flop shifts.
The shift of the walk considered here is moving shift.
On the other hand, the corresponding flip-flop type is defined by
\begin{align*}
P=\ket{\leftarrow}\bra{\rightarrow}C,\quad
R=\ket{\bullet}\bra{\bullet}C,\quad
Q=\ket{\rightarrow}\bra{\leftarrow}C.
\end{align*}
In this type, it is easily derived from a similar method that
\begin{align*}
T_N\ ({\rm Grover}) = \begin{cases}
4\quad &(N=3)
\\
\infty \quad &(N\neq 3)
\end{cases}
,
\qquad
T_N\ ({\rm Fourier}) = \begin{cases}
12\quad &(N=3)
\\
\infty \quad &(N\neq 3)
\end{cases}
.
\end{align*}

\par
\
\par

\begin{small}
\bibliographystyle{jplain}

\end{small}

\end{document}